\documentclass{uwstat518}

\usepackage{bbm}
\usepackage{amsmath}
\usepackage{graphicx} 
\usepackage{subfigure} 
\usepackage{amssymb}
\usepackage{sidecap}

\usepackage{algorithm}
\usepackage{algorithmic}

\usepackage{natbib}
\DeclareMathOperator*{\argmin}{argmin}

\newcommand{\mb}{\mathbf}

\newcommand{\R}{\mathbb{R}}
\newcommand{\C}{\mathbb{C}}

\newtheorem{theorem}{Theorem}[section]

\newtheorem{proposition}[theorem]{Proposition}

\begin{document}
\begin{center}
\Large{Efficient Transition Probability Computation for Continuous-Time Branching Processes via Compressed Sensing}

\normalsize
By Jason Xu$^\ast$, Vladimir Minin$^{\ast, \dagger}$

\small
$^\ast$Department of Statistics, University of Washington\\
$^\dagger$Department of Biology, University of Washington  \\

\normalsize
\end{center}

\begin{abstract} 
Branching processes are a class of continuous-time Markov chains (CTMCs) with ubiquitous applications. A general difficulty in statistical inference under partially observed CTMC models arises in computing transition probabilities when the discrete state space is large or uncountable. Classical methods such as matrix exponentiation are infeasible for large or countably infinite state spaces, and sampling-based alternatives are computationally intensive, requiring a large integration step to impute over all possible hidden events. Recent work has successfully applied generating function techniques to computing transition probabilities for linear multitype branching processes. While these techniques often require significantly fewer computations than matrix exponentiation, they also become prohibitive in applications with large populations. We propose a compressed sensing framework that significantly accelerates the generating function method,  decreasing computational cost up to a logarithmic factor by only assuming the probability mass of transitions is sparse. We demonstrate accurate and efficient transition probability computations in branching process models for hematopoiesis and transposable element evolution.
\end{abstract} 

\section{Introduction}
\label{Intro}
Continuous-time branching processes are widely used in stochastic modeling of population dynamics, with applications including biology, genetics, epidemiology, quantum optics, and nuclear fission \citep{renshaw2011}. With the exception of the well-studied class of birth-death processes, which have known expressions for many quantities relevant to probabilistic inference \citep{crawford2013}, branching processes pose significant inferential challenges. In particular, closed forms for finite-time \textit{transition probabilities}, the conditional probability that a trajectory ends at a given state, given a starting state and time interval, are unavailable. These transition probabilities are crucial in many inferential approaches, comprising the observed likelihood function when data from the process are available at a set of discrete times. The likelihood function is of central importance in frequentist and Bayesian methods, and any statistical framework involving observed data likelihood evaluation requires transition probability computations. Without the ability to fully leverage the branching structure, studies must rely on general CTMC estimation techniques or model approximations \citep{rosenberg2003, golinelli2006, elhay2006}. 

Computation of transition probabilities is the usual bottleneck in model-based inference using CTMCs \citep{hajiaghayi2014}, requiring a marginalization over the infinite set of possible end-point conditioned paths. Classically, this marginalization is accomplished by computing the matrix exponential of the infinitesimal generator of the CTMC. However, this procedure has cubic runtime complexity in the size of the state space, becoming prohibitive even for state spaces of moderate sizes. Alternatives also have their shortcomings: \textit{uniformization} methods use a discrete-time ``skeleton" chain to approximate the CTMC but rely on a restrictive assumption that there is a uniform bound on all rates \citep{grassmann1977, rao2011}. Typically, practitioners resort to sampling-based approaches via Markov chain Monte Carlo (MCMC). 
Specifically, particle-based methods such as sequential Monte Carlo (SMC) and particle MCMC \citep{doucet2000, andrieu2010} offer a complementary approach whose runtime depends on the number of imputed transitions rather than the size of the state space. However, these SMC methods have several limitations--- in many applications, a prohibitively large number of particles is required to impute waiting times and events between transitions, and degeneracy issues are a common occurrence, especially in longer time series. A method by \cite{hajiaghayi2014} accelerates particle-based methods by marginalizing holding times analytically, but has cubic runtime complexity in the number of imputed jumps between observations and is recommended for applications with fewer than one thousand events occurring between observations.

Recent work by \cite{xu2014} has extended techniques for computing transition probabilities in birth-death models to linear multi-type branching processes. This approach involves expanding the probability generating function (PGF) of the process as a Fourier series, and applying a Riemann sum approximation to its inversion formula. This technique has been used to compute numerical transition probabilities within a maximum likelihood estimation (MLE) framework, and has also been applied within Expectation Maximization (EM) algorithms \citep{doss2013, xu2014}. While this method provides a powerful alternative to simulation and avoids costly matrix operations, the Riemann approximation to the Fourier inversion formula requires $\mathcal{O}(N^b)$ PGF evaluations, where $b$ is the number of particle types and $N$ is the largest population size at endpoints of desired transition probabilities. This complexity is no worse than linear in the size of the state space,  but can also be restrictive: a two-type process in which each population can take values in the thousands would require millions of PGF evaluations to produce transition probabilities over an observation interval. This can amount to hours of computation in standard computing architectures, because evaluating PGFs for multitype branching processes involves numerically solving systems of ordinary differential equations (ODEs). Such computations become infeasible within iterative algorithms.

In this paper, we focus our attention on the efficient computation of transition probabilities in the presence of sparsity, presenting a novel compressed sensing generating function (CSGF) algorithm that dramatically reduces the computational cost of inverting the PGF. We apply our algorithm to a branching process model used to study hematopoiesis as well as a birth-death-shift process with applications to molecular epidemiology, and see that the sparsity assumption is valid for scientifically realistic rates of the processes obtained in previous statistical studies. We compare performance of CSGF to transition probability computations without taking advantage of sparsity, demonstrating a high degree of accuracy while achieving significant improvements in runtime.

\section{Markov Branching Processes}

A linear multitype branching process follows a population of independently acting particles that reproduce and die. The random vector $\mb{X}(t)$ takes values in a discrete state space $\Omega$ at time $t$, with $X_i(t)$ denoting the number of type $i$ particles present at time $t$. For exposition and notational simplicity, we will focus on the two-type case. In the continuous-time setting, each type $i$ particle produces $k$ type 1 particles and $l$ type 2 particles with \textit{instantaneous rates} $a_{j}(k,l)$, and the rates of no event occurring are defined as
$$ \alpha_1 := a_1(1,0) = - \sum_{(k,l) \neq (1,0)} a_1(k,l),$$  
$$ \alpha_2 := a_2(0,1) = - \sum_{(k,l) \neq (0,1)} a_2(k,l)$$
so that $\sum_{k,l} a_i(k,l) = 0$ for $i=1,2$. 	Offspring of each particle evolve according to the same set of instantaneous rates, and these rates $a_j(k,l)$ do not depend on $t$ so that the process is \textit{time-homogeneous}. These assumptions imply that each type $i$ particle has exponentially distributed lifespan with rate $-\alpha_i$, and $\mb{X}(t)$ evolves over time as a CTMC \citep{guttorp1995}.

\subsection{Transition probabilities}
Dynamics of a CTMC are determined by its transition function
\begin{equation}\label{eq:transition}
p_{\mb{x,y}}(t) = \text{Pr}(\mb{X}(t+s) = \mb{y} | \mb{X}(s) = \mb{x}), \end{equation}
where time-homogeneity implies independence of the value of $s$ on the right hand side. When the state space $\Omega$ is small, one can exponentiate the $|\Omega|$ by $|\Omega|$ \textit{infinitesimal generator} or rate matrix $\mb{Q} = \big\{ q_{\mb{x,y}} \big\}_{{\mb{x, y}} \in \Omega}$, where the entries $q_{\mb{x, y}}$ denote the instantaneous rates of jumping from state $\mb{x}$ to $\mb{y}$, to compute transition probabilities:
\begin{equation}
\mb{P}(t) := \big\{p_{\mb{x,y}}(t) \big\}_{{\mb{x, y}} \in \Omega} = e^{\mb{Q}t}  = \sum_{k=0}^\infty \frac{(\mb{Q}t)^k}{k!}.
\end{equation} 
These transition probabilities are fundamental quantities in statistical inference for data generated from CTMCs. For instance, if $\mb{X}(t)$ is observed at times $t_1, \ldots, t_J$ and $\mb{D}$ represents the $2$ by $J$ matrix containing the observed data, the \textit{observed log-likelihood} is given by
\begin{equation}\label{eq:oll} 
 \ell_o(\mb{D} ; \boldsymbol\theta ) =  \sum_{j = 1}^{J-1} \log p_{\mathbf{X}( t_j), \mathbf{X} ( t_{j+1})} ( t_{ j+1} - t_{j} ; \boldsymbol{\boldsymbol\theta} ) 
 \end{equation}
where the vector $\boldsymbol{\boldsymbol\theta}$ parametrizes the rates $a_{j}(k,l)$. Maximum likelihood inference that seeks to find the value $\hat{\boldsymbol{\boldsymbol\theta}}$ that optimizes  \eqref{eq:oll} as well as Bayesian methods where likelihood calculations arise in working with the posterior density (up to a proportionality constant) fundamentally rely on the ability to calculate transition probabilities. Having established their importance in probabilistic inference, we focus our discussion in this paper to computing these transition probabilities in a continuous-time branching process.
\subsection{Generating function methods}
Matrix exponentiation is cubic in $|\Omega|$ and thus prohibitive in many applications, but we may take an alternate approach by exploiting properties of the branching process.  \cite{xu2014} extend a generating function technique used to compute transition probabilities in birth-death processes to the multi-type branching process setting. The probability generating function (PGF) for a two-type process is defined
\begin{align}
\phi_{jk}(t, s_1, s_2 ; \boldsymbol\theta) &= \text{E}_{\boldsymbol\theta} (s_1^{X_1(t)} s_2^{X_2(t)} | X_1(0) = j, X_2(0) = k) \nonumber \\
& = \sum_{l=0}^\infty \sum_{m=0}^\infty p_{(jk),(lm)} (t ; \boldsymbol\theta) s_1^l s_2^m; \label{eq:probgen}
\end{align} this definition extends analogously for any $m$-type process. We suppress dependence on $\boldsymbol\theta$ for notational convenience.
\cite{bailey1964} provides a general technique to derive a system of differential equations governing $\phi_{jk}$ using the Kolmogorov forward or backward equations given the instantaneous rates $a_j(k,l)$. It is often possible to solve these systems analytically for $\phi_{jk}$, and even when closed forms are unavailable, numerical solutions can be efficiently obtained using standard algorithms such as Runge-Kutta methods \citep{butcher1987}. 

With $\phi_{jk}$ available, transition probabilities are related to the PGF \eqref{eq:probgen} via differentiation:
\begin{equation}\label{eq:diff}
p_{(jk),(lm)}(t) = \frac{\partial^l}{\partial s_1} \frac{\partial^m}{\partial s_2} \phi_{jk}(t) \bigg|_{s_1=s_2=0} .
\end{equation}
This repeated differentiation is computationally intensive and numerically unstable for large $l,m$, but following \cite{lange1982}, we can map the domain $s_1, s_2 \in [0,1] \times [0,1]$ to the boundary of the complex unit circle, instead setting $s_1 = e^{2 \pi i w_1}, s_2 = e^{2 \pi i w_2}$. The generating function becomes a Fourier series whose coefficients are the desired transition probabilities 
\[\phi_{jk}(t, e^{2 \pi i w_1}, e^{2 \pi i w_2} )= \sum_{l,m = 0}^\infty p_{(jk),(lm)}(t) e^{2 \pi i l w_1} e^{2 \pi i m w_2}  \]
Applying a Riemann sum approximation to the Fourier inversion formula, we can now compute the transition probabilities via integration instead of differentiation:
\begin{equation}\label{eq:FFT}
\begin{split}
 p_{(jk),(lm)} (t) &= \int_0^1 \int_0^1 \phi_{jk}(t, e^{2 \pi i w_1}, e^{2 \pi i w_2}) e^{-2 \pi i l w_1} 
 \\& \hspace{45pt} \times e^{-2 \pi i m w_2} dw_1 dw_2  \\
 & \approx  \frac{1}{N^2} \sum_{u = 0}^{N-1} \sum_{v= 0}^{N-1} \phi_{jk}(t, e^{2 \pi i u/N}, e^{2 \pi i v/N}) \\
 & \hspace{45pt} \times e^{-2 \pi i l u/N} e^{-2 \pi i m v/N} .
 \end{split}
 \end{equation}

In practice, the set of transition probabilities $S =\{ p_{(jk),(lm)}(t) \}$ for all $l,m = 0, \ldots, N$, given initial values of $(j,k)$, can be obtained via the Fast Fourier Transform (FFT), described in Section \ref{sec:CSGF}. It is necessary to choose $N > l,m$, since exponentiating the roots of unity can yield at most N distinct values $$e^{-2 \pi i m v/N} = e^{-2 \pi i ( m v \text{ mod} N)/N};$$ this is related to the Shannon-Nyquist criterion \citep{shannon2001}, which dictates that the number of samples required to recover a signal must match its highest frequency. Thus, calculating ``high frequency" coefficients--- when $l,m$ take large values---requires $\mathcal{O}(N^2)$ numerical ODE solutions, which becomes computationally expensive for large $N$. 

\paragraph{Sparsity:}
Given an initial state $\mb{X}(0) = (j,k)$, the support of transition probabilities is often concentrated over a small range of $(l,m)$ values. For example, if $\mb{X}(t) = (800,800)$, then the probability that the entire process becomes extinct, $\mb{X}(t+s) = (0,0)$, is effectively zero unless particle death rates are very high or $s$ is a very long time interval. In many realistic applications, $p_{(800,800),(l,m)}(s)$ has non-negligible mass on a small support, for instance only over $l,m$ values between $770$ and $820$. While their values can be computed using Equation \eqref{eq:FFT} for a choice of $N > 820$, requiring $N^2$ ODE evaluations toward computing only $(820-770)^2$ nonzero probabilities seems wasteful.  To exploit the sparsity information in such a setting, we bridge aforementioned branching process techniques to the literature of \textit{compressed sensing}.

\section{Compressed Sensing}
Originally developed in an information theoretic setting, the principle of compressed sensing (CS) states that an unknown sparse signal can be recovered accurately and often perfectly from significantly fewer samples than dictated by the Shannon-Nyquist rate at the cost of solving a convex optimization problem \citep{donoho2006, candes2006}. CS is a robust tool to collect high-dimensional sparse data from a low-dimensional set of measurements and has been applied to a plethora of fields, leading to dramatic reductions in the necessary number of measurements, samples, or computations. In our setting, the transition probabilities play the role of a target sparse signal of Fourier coefficients. The data reduction made possible via CS then translates to reducing computations to a random subsample of PGF evaluations, which play the role of measurements used to recover the signal.

\subsection{Overview}\label{sec:overview}
In the CS framework, the unknown signal is a vector $\mb{x} \in \C^N$ observed through a measurement $\mb{b} = \mb{V} \mb{x}~\in~\C^M$ with $M < < N$. Here $\mb{V}$ denotes an $M \times N$ \textit{measurement matrix} or sensing matrix. Since $M < N$, the system is underdetermined and inversion is highly ill-posed---the space of solutions is an infinite affine subspace, but CS theory shows that recovery can be accomplished under certain assumptions by seeking the \textit{sparsest} solution. Let $\boldsymbol\psi$ be an orthonormal basis of $\C^N$ that allows a $K$-sparse representation of $\mb{x}$: that is, $\mb{x} = \boldsymbol\psi \mb{s}$ where $\mb{s}$ is a sparse vector of coefficients such that $||\mb{s}||_0 < K$. \cite{candes2006} proves that recovery can then be accurately accomplished by finding the sparsest solution 
\begin{equation}\label{eq:l0}
 \hat{\mb{s}} = \argmin _\mb{s} ||\mb{s}||_0 \hspace{12pt} s.t. \hspace{12pt} \mb{A}\mb{s} = \mb{b}
 \end{equation} 
 where $\mb{A} = \mb{V} \boldsymbol\psi$ is the composition of the measurement and sparsifying matrices.
In practice, this non-convex objective is combinatorially intractable to solve exactly, and is instead solved by proxy via $\ell_1$-relaxation, resulting in a convex optimization program. In place of Equation $\eqref{eq:l0}$, we optimize the unconstrained penalized objective \begin{equation}\label{eq:l1}
 \hat{\mb{s}}  = \argmin _{\mb{s}} \frac{1}{2}|| \mb{A} \mb{s} - \mb{b}||_2^2 + \lambda ||\mb{s}||_1 
 \end{equation} where $\lambda$ is a regularization parameter enforcing sparsity of $\mb{s}$.
The signal $\mb{x}$, or equivalently $\mb{s}$, can be recovered perfectly using only $M = C K \log N$ measurements for some constant $C$ when $\mb{A}$ satisfies the \textit{Restricted Isometry Property} (RIP) \citep{candes2005, candes2008}---briefly, this requires that $\mathbf{V}$ and $\boldsymbol \psi$ to be \textit{incoherent} so that rows of  $\mathbf{V}$ cannot sparsely represent the columns of $\boldsymbol\psi$ and vice versa. Coherence between $\mathbf{V}, \boldsymbol\psi$ is defined as
\[ \mu( \mathbf{V}, \boldsymbol\psi) = \sqrt{n} \max_{i,j} | \langle \mathbf{V}, \boldsymbol\psi_j \rangle |, \]
and low coherence pairs are desirable. It has been shown that choosing random measurements $\mathbf{V}$ satisfies RIP with overwhelming probability \citep{candes2008}. Further, given $\boldsymbol \psi$, it is often possible to choose a known ideal distribution from which to sample elements in $\mathbf{V}$ such that $\mathbf{V}$ and $\boldsymbol{\psi}$ are maximally incoherent.

\subsection{Higher dimensions}
CS theory extends naturally to higher-dimensional signals \citep{candes2006}. In the 2D case which will arise in our applications (Section \ref{sec:examples}), the sparse solution $\mathbf{S} \in \C^{N \times N}$ and measurement 
\begin{equation}\label{eq:measurement}
\mb{B} = \mb{A S A} ^T \in \C^{M \times M} \end{equation}
are matrices rather than vectors, and we solve
\begin{equation}\label{eq:2d}
 \hat {\mb{S}} = \argmin _{\mb{S}} \frac{1}{2}|| \mb{A S A}^T -\mb{B} ||_2^2 + \lambda ||\mb{S}||_1 .
 \end{equation} 
 This can always be equivalently represented in the vector-valued framework: vectorizing 
 $$\text{vec}(\mb{S}) = \widetilde{\mb{s}} \in \C^{N^2}, \quad \text{vec}(\mb{B}) = \widetilde{\mb{b}} \in \C^{M^2},$$ 
we now seek $\widetilde{\mb{b}} = \widetilde{A} \widetilde{\mb{s}}$ as in Equations \eqref{eq:l0}, \eqref{eq:l1}, where $\widetilde{\mb{A}} = \mb{A} \otimes \mb{A}$ is the Kronecker product of $\mb{A}$ with itself. In practice, it can be preferable to solve \eqref{eq:2d}, since the number of entries in $\widetilde{\mb{A}}$ grows rapidly and thus the vectorized problem requires a costly construction of $\widetilde{\mb{A}}$ and can be cumbersome in terms of memory. 

\section{CSGF Method}\label{sec:CSGF}
We propose an algorithm that allows for efficient PGF inversion within a compressed sensing framework. We focus our exposition on two-type models: linear complexity in $|\Omega|$ is less often a bottleneck in single-type problems, and all generating function methods as well as compressed sensing techniques we describe extend to higher dimensional settings.

We wish to compute the transition probabilities $p_{jk,lm}(t)$ given any $t>0$ and $\mb{X}(0) = (j,k)$. These probabilities can be arranged in a matrix $\mb{S} \in \R^{N\times N} $ with entries 
$$\big\{ \mb{S} \big\}_{l,m} = p_{jk,lm}(t).$$ 
Without the CS framework, these probabilities are obtained following Equation \eqref{eq:FFT} by first computing an equally sized matrix of PGF solutions 
\begin{equation}\label{eq:full}
\widetilde{ \mb{B} } = \big\{ \ \phi_{jk}(t, e^{2 \pi i u/N}, e^{2 \pi i v/N}) \big\}_{u,v = 0}^{N-1} \in \C^{N \times N}. \end{equation}
 For large $N$, obtaining $\widetilde{\mb{B}}$ is computationally expensive, and our method seeks to bypass this step. When $\widetilde{\mb{B}}$ is computed, transition probabilities are then recovered by taking the fast Fourier transform $\mb{S} = \text{fft}(\widetilde{\mb{B}})$. To better understand how this fits into the CS framework, we can equivalently write the fast Fourier transform in terms of matrix operations $ \mb{S} = \mb{ F \widetilde{B} F }^T$, where $\mb{F} \in \C^{N \times N}$ denotes the discrete Fourier transform matrix (see Supplement). Thus, the sparsifying basis $\boldsymbol\psi$ is the Inverse Discrete Fourier Transform (IDFT) matrix $\boldsymbol\psi = \mb{F}^*$ given by the conjugate transpose of $\mb{F}$, and we have $\widetilde{\mb{B}} = \boldsymbol\psi \mb{S} \boldsymbol \psi^T$.

When the solution matrix $\mb{S}$ is expected to have a sparse representation, our CSGF method seeks to recover $\mb{S}$ without computing the full matrix $\widetilde{\mb{B}}$, instead beginning with a much smaller set of PGF evaluations $\mb{B} \in \C^{M \times M}$ corresponding to random entries of $\widetilde{\mb{B}}$ selected uniformly at random. Denoting randomly sampled indices $\mathcal{I}$, this smaller matrix is a projection $\mb{B} = \mb{A S A}^T$ in the form of Equation \eqref{eq:measurement} where $\mb{A} \in \C^{M\times N}$ is obtained by selecting a subset of rows of $\boldsymbol \psi$ corresponding to $\mathcal{I}$. Uniform sampling of rows corresponds to multiplying by a measurement matrix encoding the \textit{spike basis} (or standard basis): formally, this fits into the framework described in Section \ref{sec:overview} as $\mb{A} = \mb{V} \boldsymbol\psi$, with measurement matrix rows $\mb{V}_j(l) = \delta(j-l)$. The spike and Fourier bases are known to be \textit{maximally incoherent} in any dimension, so uniformly sampling indices $\mathcal{I}$ is optimal in our setting. 

Now in the compressed sensing framework, computing the reduced matrix $\mb{B}$ only  requires a logarithmic proportion $|\mb{B}| \propto K \log |\widetilde{\mb{B}}|$ of PGF evaluations necessary in Equation \eqref{eq:full}. Computing transition probabilities in $\mb{S}$ is thus reduced to a signal recovery problem, solved by optimizing the objective in Equation \eqref{eq:2d}.

\subsection{Solving the $\ell_1$ problem}
There has been extensive research on algorithms for solving the $\ell_1$ regularization objective in Equation \eqref{eq:l1} and related problems \citep{tibshirani1996, beck2009}. As mentioned previously, vectorizing the problem so that it can be represented in the form $\eqref{eq:l1}$ requires wasteful extra memory; instead we choose to solve the objective in Equation \eqref{eq:2d} using a \textit{proximal gradient descent} (PGD) algorithm. 

PGD is useful for solving minimization problems with objective of the form $f(x) = g(x) + h(x)$ with $g$ convex and differentiable, and $h$ convex but not necessarily differentiable. Letting $$g(\mb{S}) =  \frac{1}{2}||\mb{ASA}^T- \mb{B}||_2^2, \quad h(\mb{S}) = \lambda ||\mb{S}||_1,$$
we see that Equation \eqref{eq:2d} satisfies these conditions. 
A form of generalized gradient descent, PGD iterates toward a solution with
\begin{align}\label{eq:prox}
x_{k+1} = \argmin_z [ g(x_k) &+ \nabla g(x_k)^T (z - x_k)\\ \nonumber
& + \frac{1}{2 L_k} || z - x_k ||_2^2 + h(z) ],
\end{align}
where $L_k$ is a step size that is either fixed or determined via line-search. This minimization has known closed-form solution
\begin{equation}\label{eq:proxsol}
x_{k+1} = \text{softh} ( x_k - L_k \nabla g(x_k), L_k \lambda ), \end{equation}
where \text{softh} is the soft-thresholding operator 
\begin{equation}\label{eq:softmax}
[ \text{softh}(x, \alpha) ]_i = \text{sgn}(x_i) \max (|x_i| - \alpha, 0) .  \end{equation}
Alternating between these steps results in an \textit{iterative soft-thresholding algorithm} that solves the convex problem $\eqref{eq:2d}$ with rate of convergence $\mathcal{O}(1/k)$ when $L_k$ is fixed. 
The \text{softh}$()$ operation is simple and computationally negligible, so that the main computational cost is in evaluating $\nabla g(x_k)$. We derive a closed form expression for the gradient in our setting
\begin{equation}\label{eq:gradient}
 \nabla g(\mb{S}) = - \mb{A}^* (\mb{B} - \mb{A S A}^T) \overline{\mb{A}} , \end{equation}
where $\overline{\mb{A}}, \mb{A}^*$ denote complex conjugate and conjugate transpose of $\mb{A}$ respectively. In practice, the inner term $\mb{ASA}^T$ is obtained as a subset of the inverse fast Fourier transform of $\mb{S}$ rather than by explicit matrix multiplication. The computational effort in computing $\nabla g(\mb{S})$ therefore involves only the two outer matrix multiplications.

We implement a fast variant of PGD using momentum terms \citep{beck2009fast} based on an algorithm introduced by Nesterov, and select step sizes $L_k$ via a simple line-search subroutine \citep{beck2009}. The accelerated version includes an \textit{extrapolation step}, where the soft-thresholding operator is applied to a momentum term $$y_{k+1} = x_{k}+ \omega_k(x_k- x_{k-1})$$ rather than to $x_k$; here $\omega_k$ is an extrapolation parameter for the momentum term.
Remarkably, the accelerated method still only requires one gradient evaluation at each step as $y_{k+1}$ is a simple linear combination of previously computed points, and has been proven to achieve the optimal worst-case rate of convergence $\mathcal{O}(1/k^2)$ among first order methods \citep{nesterov1983}. Similarly, the line-search procedure involves evaluating a bound that also only requires one evaluation of $\nabla g$ (see Supplement).

Algorithm \ref{alg:CSGF} provides a summary of the CSGF method in pseudocode.

\begin{algorithm}[h!]
   \caption{\texttt{CSGF} algorithm.}
   \label{alg:CSGF}
   \begin{algorithmic}[1]
\STATE \textbf{Input:} initial sizes $X_1=j, X_2=k$, time interval $t$, branching rates $\boldsymbol\theta$, signal size $N > j,k$, measurement size $M$, penalization constant $\lambda > 0$, line-search parameters $L, c$.
\STATE Uniformly sample $M$ indices $\mathcal{I} ~\subset \left[0, \ldots N-1 \right]/N$
\STATE Compute $\mb{B} = \big\{  \ \phi_{jk}(t, e^{2 \pi i u/N}, e^{2 \pi i v/N}) \big\}_{u,v \in \mathcal{I} \times \mathcal{I}} $
\STATE Define $\mb{A} = \boldsymbol\psi_{\mathcal{I} \cdot}$ the $\mathcal{I}$ rows of IDFT matrix $\boldsymbol{\psi}$
\STATE \textbf{Initialize:} $\mb{S}_1 = \mb{Y}_1 = \mb{0}$
\FOR {$k = 1, 2, \ldots, \left\{ \text{max iterations} \right\} $ }
\STATE Choose $L_{k} = \texttt{line-search}(L, c, \mb{Y}_{k} )$
\STATE Update extrapolation parameter $\omega_{k} = \frac{k}{k+3}$
\STATE Update momentum $\mb{Y}_{k+1} = \mb{S}_k + \omega_k( \mb{S}_{k} - \mb{S}_{k-1})$
\STATE Compute $\nabla g(\mb{Y}_{k+1})$ according to \eqref{eq:gradient}
\STATE Update $\mb{S}_{k+1} = \text{softh}(\mb{S}_k - L_k \nabla g(\mb{Y}_{k+1}), L_k \lambda)$
\ENDFOR
\RETURN $\hat{\mb{S}} = \mb{S}_{k+1}$
\end{algorithmic}
\end{algorithm}

\section{Examples}\label{sec:examples}
We will examine the performance of CSGF in two applications: a stochastic two-compartment model used in statistical studies of \textit{hematopoiesis}, the process of blood cell production, and a birth-death-shift model that has been used to study the evolution of \textit{transposons}, mobile genetic elements.

\subsection{Two-compartment hematopoiesis model}
\begin{SCfigure}
\centering
\vspace{1pt}
\includegraphics[width = .26\paperwidth]{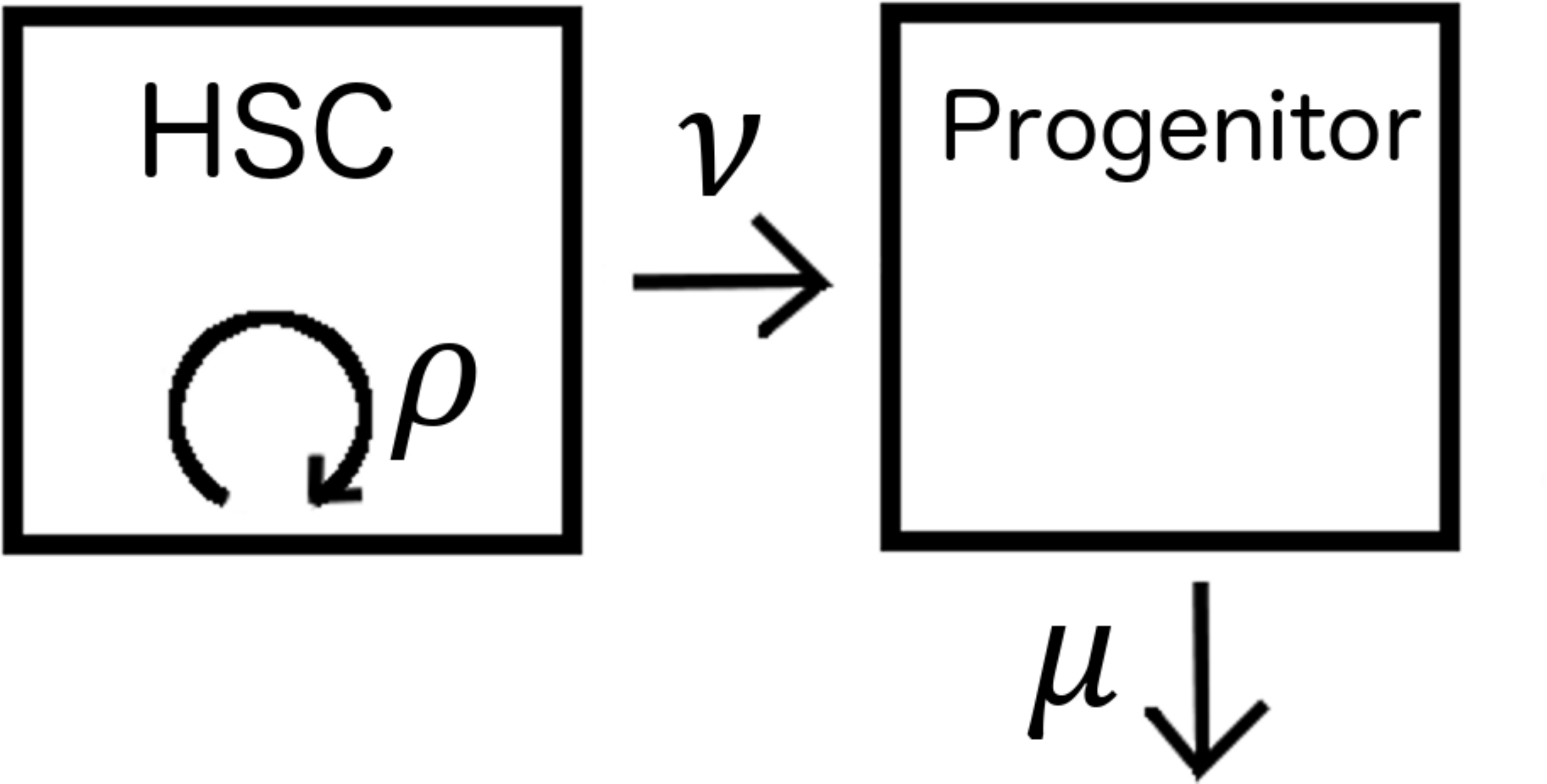}
\caption{HSCs can self-renew, producing new HSCs at rate $\rho$, or differentiate into progenitor cells at rate $\nu$. Further progenitor differentiation is modeled by rate $\mu$. }
\label{fig:HSC}
\end{SCfigure}

 Hematopoiesis is the process in which self-sustaining primitive hematopoietic stem cells (HSCs) specialize, or \textit{differentiate}, into progenitor cells, which further specialize to eventually produce mature blood cells. In addition to far-reaching clinical implications --- stem cell transplantation is a mainstay of cancer therapy --- understanding hematopoietic dynamics is biologically interesting, and provides critical insights of general relevance to other areas of stem cell biology \citep{orkin2008}. The stochastic model, depicted in Figure \ref{fig:HSC}, has enabled estimation of hematopoietic rates in mammals from data in several studies \citep{catlin2001, golinelli2006, fong2009}. Without the ability to compute transition probabilities, an estimating equation approach by \cite{catlin2001} is statistically inefficient, resulting in uncertain estimated parameters with very wide confidence intervals. Nonetheless, biologically sensible rates are inferred.
\cite{golinelli2006} observe that transition probabilities are unknown for a linear birth-death process (compartment 1) coupled with an inhomogeneous immigration-death process (compartment 2), motivating their computationally intensive reversible jump MCMC implementation. 

However, we can equivalently view the model as a two-type branching process. Under such a representation, it becomes possible to compute transition probabilities via Equation \eqref{eq:FFT}. The type one particle population $X_1$ corresponds to hematopoietic stem cells (HSCs), and $X_2$ represents progenitor cells. With parameters as denoted in Figure \ref{fig:HSC}, the nonzero instantaneous rates defining the process are 
\begin{align}
a_1(2,0) &= \rho && a_1(0,1) = \nu && a_1(1,0) = -(\rho + \nu) \nonumber \\
a_2(0,0) &= \mu && a_2(0,1) = -\mu. \label{eq:HSC}
\end{align}

Having specified the two-type branching process, we derive solutions for its PGF, defined in Equation \eqref{eq:probgen}, with details in the Supplement:
\begin{proposition}
The generating function for the two-type model described in \eqref{eq:HSC} is given by $\phi_{jk}~=~\phi_{1,0}^j \phi_{0,1}^k$, where
\begin{equation} \label{eq:HSCODE} 
\begin{cases}
\phi_{0,1}(t, s_1, s_2) = 1 + (s_2 - 1)e^{-\mu t} \\
\frac{d}{dt} \phi_{1,0}(t, s_1, s_2) = \rho \phi_{1,0}^2(t, s_1, s_2) - (\rho + \nu) \phi_{1,0}(t,s_1, s_2) \\
\hspace{70pt} + \nu \phi_{0,1}(t, s_1, s_2).
\end{cases}
\end{equation}
\end{proposition}
We see that $\phi_{0,1}$ has closed form solution so that evaluating $\phi_{jk}$ only requires solving one ODE numerically, and with the ability to compute $\phi_{jk}$, we may obtain transition probabilities using Equation \eqref{eq:FFT}. In this application, cell populations can easily reach thousands, motivating the CSGF approach to accelerate transition probability computations.

\subsection{Birth-death-shift model for transposons}
Our second application examines the birth-death-shift (BDS) process proposed by \cite{rosenberg2003} to model evolutionary dynamics of transposable elements or \textit{transposons}, genomic mobile sequence elements. Each transposon can (1) duplicate, with the new copy moving to a new genomic location; (2) shift to a different location; or (3) be removed and lost from the genome, independently of all other transposons. These respective birth, shift, and death events occur at per-particle instantaneous rates $\beta, \sigma, \delta$, with overall rates proportional to the total number of transposons. Transposons thus evolve according to a linear \textit{birth-death-shift} Markov process in continuous time. In practice, genotyping technologies allow for this process to be discretely monitored, necessitating computation of finite-time transition probabilities.

\cite{rosenberg2003} estimate evolutionary rates of the IS\textit{6110} transposon in the \textit{Mycobacterium tuberculosis} genome from a San Francisco community study dataset \citep{cattamanchi2006}. Without transition probabilities, the authors maximize an approximate likelihood by assuming at most one event occurs per observation interval, a rigid assumption that severely limits the range of applications. \cite{doss2013} revisit their application, inferring similar rates of IS\textit{6110} evolution using a one-dimensional birth-death model that ignores shift events. \cite{xu2014} show that the BDS model over any finite observation interval can be modeled as a two-type branching process, where $X_1$ denotes the number of initially occupied genomic locations and $X_2$ denotes the number of newly occupied locations (see figure in Supplement). In this representation, full dynamics of the BDS model can be captured, and generating function techniques admit transition probabilities, leading to rate estimation via MLE and EM algorithms.  Transposon counts in the tuberculosis dataset are low, so that Equation \eqref{eq:FFT} can be computed easily, but their method does not scale well to applications with high counts in the data.

The nonzero rates defining the two-type branching process representation of the BDS model are given by
\begin{align}
a_1(1,1) &= \beta, && a_1(0,1) = \sigma, && a_1(0,0) = \delta, \nonumber \\
 a_1(1,0) &= -(\beta + \sigma + \delta), && a_2(0,2) = \beta, \nonumber \\
 a_2(0,1) &= -(\beta + \delta), && a_2(0,0) = \delta.  \label{eq:rates}
\end{align}
and its PGF is governed by the following system derived in \citep{xu2014}:
\begin{equation}\label{eq:BDSODE}
\begin{cases}
\phi_{0,1}(t,s_1,s_2) = 1 + \left[ \frac{\beta}{\delta - \beta} + (\frac{1}{s_2-1} + \frac{\beta}{\beta - \delta})e^{(\delta - \beta)t} \right] ^{-1} \\
 \frac{d}{dt} \phi_{1,0}(t,s_1,s_2) = \beta \phi_{1,0} \phi_2 + \sigma  \phi_{0,1} + \delta - (\beta + \sigma + \delta) s_1,
\end{cases}
\end{equation}
again with $\phi_{jk} = \phi_{1,0}^j \phi_{0,1}^k$ by particle independence.

\subsection{Results}

\begin{figure}
\vspace{-70pt}
\includegraphics[width = .65\paperwidth]{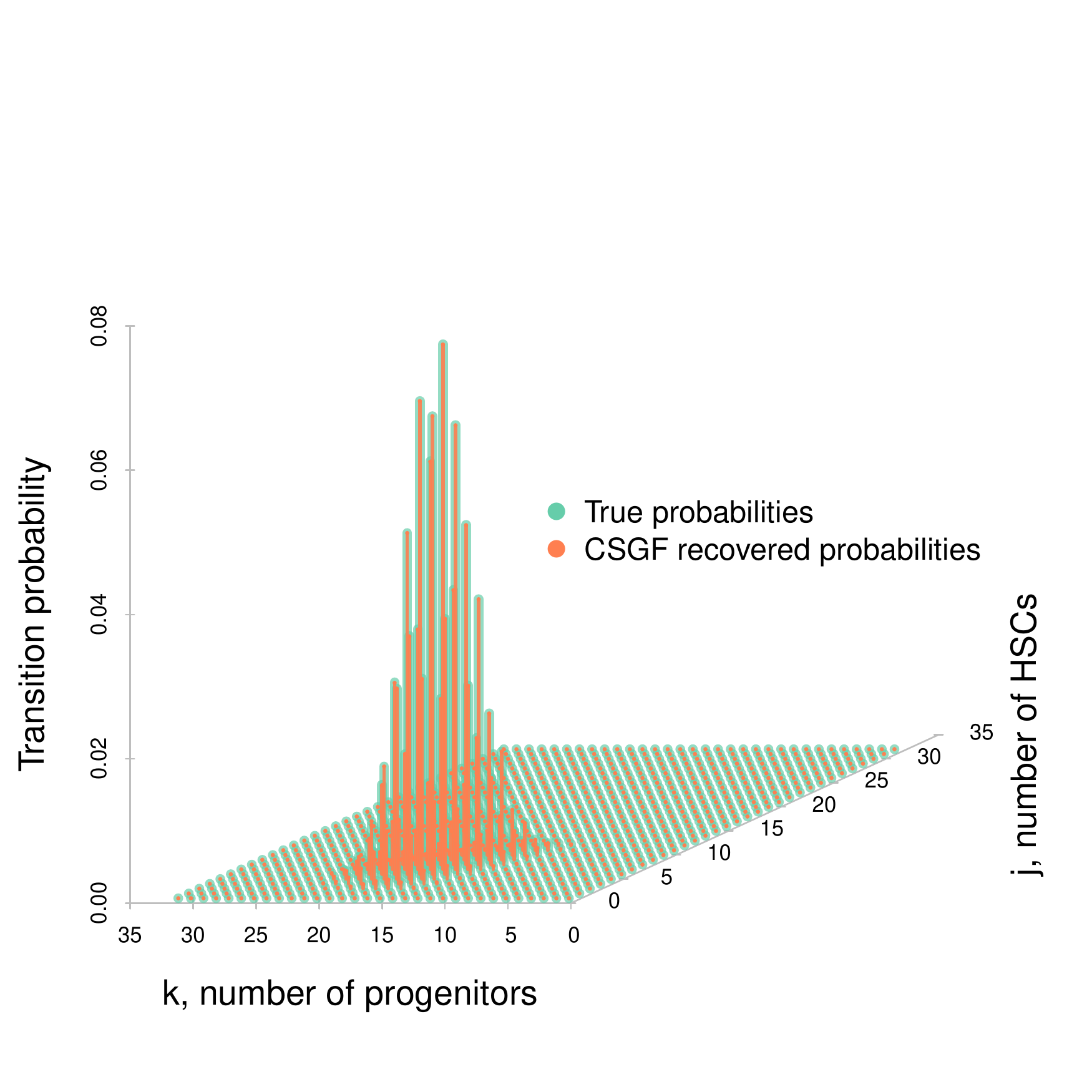}
\centering
\vspace{-25pt}
\caption{Illustrative example of recovered transition probabilities in hematopoiesis model described in Section \ref{sec:examples}. Beginning with 15 HSCs and 5 progenitors over a time period of one week, the CSGF solution $\hat{\mb{S}} = \big\{ \hat p_{(15,5),(j,k)}(1) \big\}$, $j, k = 0,\ldots,31$, perfectly recovers transition probabilities $\mb{S}$, using fewer than half the measurements.} 
\label{fig:HSCtrans}
\end{figure}

To compare the performance of CSGF to the computation of Equation \eqref{eq:FFT} without considering sparsity, we first compute sets of transition probabilities $\mb{S}$ of the 
hematopoiesis model using the full set of  PGF solution measurements $\widetilde{\mb{B}}$ as described in Equation \eqref{eq:full}. These ``true signals" are compared to the signals computed using CSGF $\hat{\mb{S}}$, recovered using only a random subset of measurements $\mb{B}$ following Algorithm \ref{alg:CSGF}. Figure \ref{fig:HSCtrans} provides an illustrative example with small cell populations for visual clarity--- we see that the support of transition probabilities is concentrated (sparse), and the set of recovered probabilities $\hat{\mb{S}}$ is visually identical to the true signal.

In each of the aforementioned applications, we calculate transition probabilities $\mb{S} \in \R^{N \times N}$ for maximum populations $N = 2^7, 2^8, \ldots 2^{12}$, given  rate parameters $\boldsymbol\theta$, initial population $\mb{X}(0)$, and time intervals $t$. Each computation of $\mb{S}$ requires $N^2$ numerical evaluations of the ODE systems \eqref{eq:HSCODE}, \eqref{eq:BDSODE}. For each value of $N$, we repeat this procedure beginning with ten randomly chosen sets of initial populations $\mb{X}(0)$ each with total size less than $N$. We compare the recovered signals $\hat{\mb{S}}$ computed using CSGF to true signals $\mb{S}$, and report median runtimes and measures of accuracy over the ten trials, with details in the following sections.

\begin{figure}[H]
\centering
\includegraphics[width = .65\paperwidth]{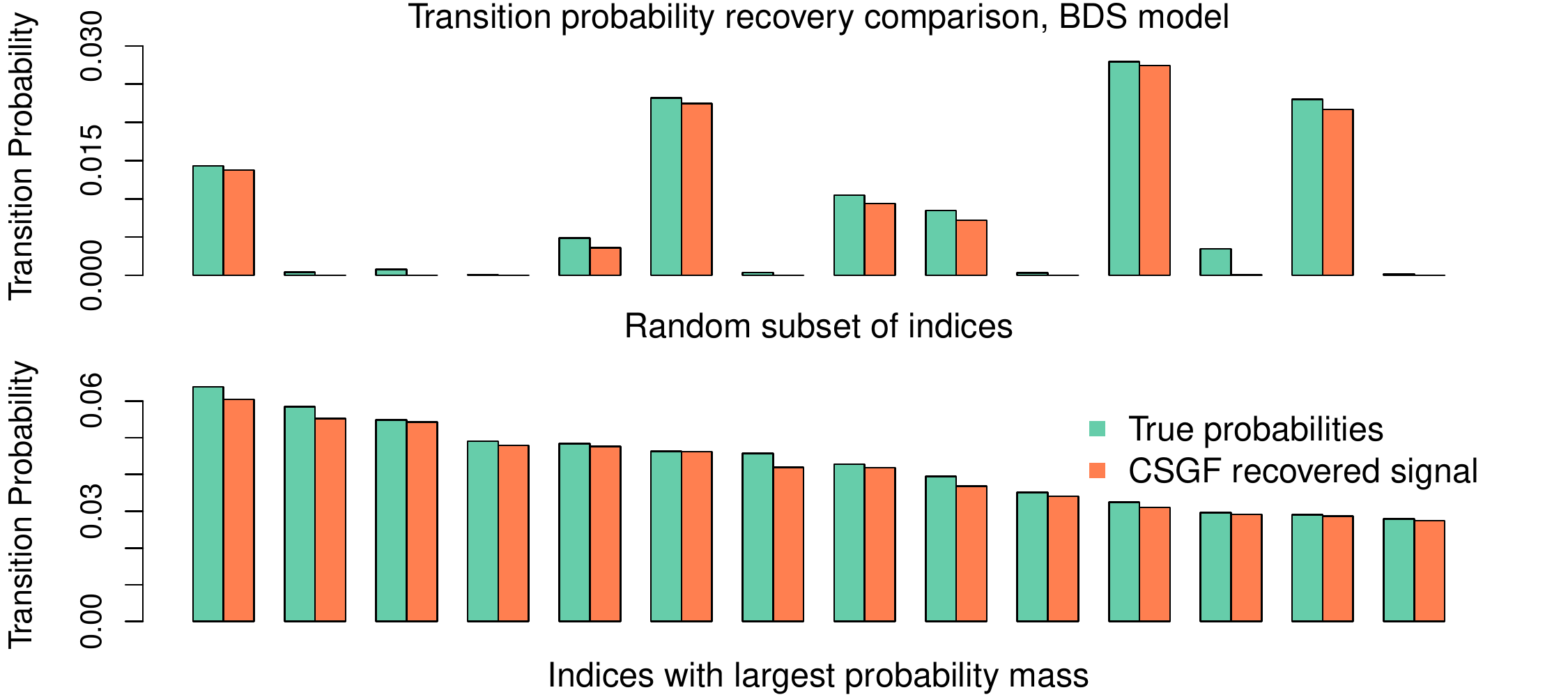} 
\caption{Randomly selected probabilities and largest probabilities recovered using CSGF are nearly identical to their true values. Probabilities displayed here correspond to a randomly selected BDS model trial with N=512; transition probabilities $\hat{\mb{S}}$ via CSGF are recovered from a sample $\mb{B}$ requiring fewer than $2\%$ of ODE computations used to compute $\mb{S} = \text{fft}(\widetilde{\mb{B}})$.} 
\label{fig:BDSprobs}
\end{figure}

\begin{table*}[!ht]
\begin{center}
\caption{Runtimes and error, birth-death-shift model. }

\begin{tabular}{|r|r|r|r|r|r|r|}
 \hline
$N$ & $M$  & \shortstack{Time (sec), \\ $\widetilde{\mb{B}} \in \C^{N \times N}$ } & \shortstack{Time (sec), \\ $\mb{B} \in \C^{M \times M}$ }  & \shortstack{Time (sec), \\ PGD}  & \shortstack{$\varepsilon_\text{max} = $ \\ $|\hat{p}_{ij,kl} -p_{ij,kl}|_\text{max}$ } & \shortstack{ $\varepsilon_\text{rel} = $ \\ $\varepsilon_\text{max}/{ |p_{ij,kl} |_\text{max} }$  } \\ 
\hline
128 & 	25 & 39.7 		& 2.3		& 1.0			& $5.27 \times 10^{-3}$	& $2.77 \times 10^{-2}$ \\	
256 & 	33 & 	150.2	& 3.8		& 7.8			& $4.86 \times 10^{-3}$	& $4.71 \times 10^{-2}$\\
512 & 	45 & 	895.8	& 7.8		& 25.3		& $2.71 \times 10^{-3}$	& $4.68 \times 10^{-2}$\\
1024 & 	68 & 	2508.9	& 18.6	& 58.2		& $1.41 \times 10^{-3}$	& $5.12 \times 10^{-2}$ \\
2048 & 	101 & 9788.3	& 26.1	& 528.3		&  $8.10 \times 10^{-4}$	& $4.81 \times 10^{-2}$\\ 
4096 &   150 & 	40732.7	& 57.4	& 2234.7		& $4.01 \times 10^{-4}$	& $5.32 \times 10^{-2}$\\
\hline
\end{tabular}
\label{tab:BDS}

\caption{Runtimes and error, hematopoiesis model }
\begin{tabular}{|r|r|r|r|r|r|r|}
 \hline
$N$ & $M$  & \shortstack{Time (sec), \\ $\widetilde{\mb{B}} \in \C^{N \times N}$ } & \shortstack{Time (sec), \\ $\mb{B} \in \C^{M \times M}$ } & \shortstack{Time (sec), \\ PGD}  & \shortstack{$\varepsilon_\text{max} = $ \\ $|\hat{p}_{ij,kl} -p_{ij,kl}|_\text{max}$ } & \shortstack{ $\varepsilon_\text{rel} = $ \\ $\varepsilon_\text{max}/{ |p_{ij,kl} |_\text{max} }$  } \\ 
\hline
128 & 	43 & 108.6 	& 9.3		& 0.64		& $9.41 \times 10^{-4}$	& $2.25 \times 10^{-2}$ \\	
256 & 	65 & 	368.9	& 22.1	& 2.1			& $9.44 \times 10^{-4}$	& $4.73 \times 10^{-2}$\\
512 & 	99 & 	922.1	& 44.8	& 8.5			& $3.23 \times 10^{-4}$	& $3.60 \times 10^{-2}$\\
1024 & 	147 & 5740.1	& 118.1	& 41.9		& $2.27 \times 10^{-4}$	& $5.01 \times 10^{-2}$ \\
2048 & 	217 & 12754.8	& 145.0	& 390.0		&  $1.29 \times 10^{-4}$	& $5.10 \times 10^{-2}$\\ 
4096 &   322 & 	58797.3	& 310.7	& 2920.3		& $9.43 \times 10^{-5}$	& $6.13 \times 10^{-2}$\\
\hline
\end{tabular}
\label{tab:HSC}

\end{center}
\end{table*}

\paragraph{Parameter settings:}
In the hematopoiesis example, we set per-week rates $\boldsymbol\theta_\text{hema}~=~(0.125, 0.104, 0.147)$ and observation time $t=1$ week based on biologically sensible rates and observation time scales of data from previous studies of hematopoiesis in mammals \citep{catlin2001, golinelli2006, fong2009}. For the BDS application, we set per-year event rates  $\boldsymbol\theta_\text{bds} = (0.0156, 0.00426, 0.0187)$ estimated in \citep{xu2014}, and $t= .35$ years, the average length between observations in the San Francisco tuberculosis dataset \citep{cattamanchi2006}. 
 
In each case, we computed $M^2 = 3 K \log N^2$ total random measurements to obtain $\mb{B}$ for CSGF, and we set the regularization parameters $\lambda_\text{hsc} = \sqrt{  \log M }$, $\lambda_\text{bds} = \log M$, with more regularization in the BDS application as lower rates and a shorter observation interval leads us to expect more sparsity. While careful case-by-case tuning to choose $\lambda, M$ would lead to optimal results, we set them in this simple manner across \textit{all} trials to demonstrate a degree of robustness, still yielding promising performance results. In practice one may apply standard cross-validation procedures to select $\lambda, M$, and because the target solution is a set of transition probabilities, checking that entries in the recovered solution $\hat{\mb{S}}$ sum close to 1 offers a simpler available heuristic. Finally, though one may expedite convergence of PGD by supplying an informed initial guess with positive values near values $\mb{X}(0)$ in practice, we initialize PGD with an uninformative initial value $\mb{S}_1 = \mb{0}$ in all cases.

\paragraph{Accuracy:}
In both models and for all values of $N$, each signal was reconstructed very accurately. 
Errors are reported in Tables \ref{tab:BDS} and \ref{tab:HSC} for the BDS and hematopoiesis models respectively. Maximum absolute errors for each CSGF recovery 
$$ \varepsilon_{\max} = \max_{kl} | \{ \hat{\mb{S}} \}_{kl} - \left\{ \mb{S} \right\}_{kl} | = \max_{kl} | \hat{p}_{ij,kl}(t) -p_{ij,kl}(t) | $$  
are on the order of $10^{-3}$ at worst. We also report a measure of relative error, and because $\varepsilon_{\max}$ is typically attained at large probabilities, we include the maximum absolute error relative to the largest transition probability $$\varepsilon_\text{rel} = \frac{\varepsilon_{\max}}{ \max_{kl} \left\{ S \right\}_{kl}}, $$ providing a more conservative measure of accuracy. We still see that $\varepsilon_\text{rel}$ is on the order of $10^{-2}$ in all cases. Visually, the accuracy of CSGF is stark: Figure \ref{fig:BDSprobs} provides a side-by-side comparison of randomly selected transition probabilities recovered in the BDS model for $N=2^9$.

\paragraph{Running Times:}
Tables \ref{tab:BDS} and \ref{tab:HSC} show dramatic improvements in runtime using CSGF, reducing the number of ODE computations logarithmically. For instance, with $N=4096$, we see the time spent on PGF evaluations necessary for CSGF is less than $0.1 \%$ of the time required to compute $\mb{S}$ in the BDS model, and around $0.5\%$ of computational cost in the less sparse hematopoiesis application. Including the time required for solving Equation \eqref{eq:2d} via PGD, we see that computing $\hat{\mb{S}}$ using CSGF reduces runtime by two orders of magnitude, requiring less than $6\%$ of total computational time spent toward computing $\mb{S}$ in the worst case. We remark that ODE solutions are computed using a C implementation of efficient solvers via package \texttt{deSolve}, while we employ a naive R implementation of PGD. We emphasize the logarithmic reduction in required numerical ODE solutions; an optimized implementation of PGD reducing R overhead will yield further real-time efficiency gains.

\section{Discussion}
We have presented a novel adaptation of recent generating function techniques to compute branching process transition probabilities within the compressed sensing paradigm. While generating function approaches bypass costly matrix exponentiation and simulation-based techniques by exploiting mathematical properties in the branching structure, our contribution now makes these techniques scalable by additionally harnessing the available sparsity structure. We show that when sparsity is present in the set of transition probabilities, computational cost can be reduced up to a logarithmic factor over existing methods. Note that sparsity is the \textit{only} additional assumption necessary to apply our CSGF method---no prior knowledge about where transition probabilities have support is necessary. Many real-world applications of branching process modeling feature such sparsity, and we have seen that CSGF achieves accurate results with significant efficiency gains in two such examples with realistic parameter settings from the scientific literature. Transition probabilities are often important, interpretable quantities in their own right, and are necessary within any likelihood-based probabilistic framework for partially observed CTMCs.  Their tractability using CSGF opens doors to applying many Bayesian and frequentist tools to settings in which such methods were previously infeasible. Finally, we note that other statistically relevant quantities such as expectations of particle dwell times and restricted moments can be computed using similar generating function techniques \citep{minin2008}, and the CSGF framework applies analogously when sparsity is present.

\bibliography{CS_UAI}

\begin{thebibliography}{30}
\providecommand{\natexlab}[1]{#1}
\providecommand{\url}[1]{\texttt{#1}}
\expandafter\ifx\csname urlstyle\endcsname\relax
  \providecommand{\doi}[1]{doi: #1}\else
  \providecommand{\doi}{doi: \begingroup \urlstyle{rm}\Url}\fi

\bibitem[Andrieu et~al.(2010)Andrieu, Doucet, and Holenstein]{andrieu2010}
C~Andrieu, A~Doucet, and R~Holenstein.
\newblock Particle {M}arkov chain {M}onte {C}arlo methods.
\newblock \emph{Journal of the Royal Statistical Society: Series B (Statistical
  Methodology)}, 72\penalty0 (3):\penalty0 269--342, 2010.

\bibitem[Bailey(1964)]{bailey1964}
NTJ Bailey.
\newblock \emph{The Elements of Stochastic Processes; with Applications to the
  Natural Sciences}.
\newblock New York: Wiley, 1964.

\bibitem[Beck and Teboulle(2009{\natexlab{a}})]{beck2009}
A~Beck and M~Teboulle.
\newblock Gradient-based algorithms with applications to signal recovery.
\newblock \emph{Convex Optimization in Signal Processing and Communications},
  2009{\natexlab{a}}.

\bibitem[Beck and Teboulle(2009{\natexlab{b}})]{beck2009fast}
A~Beck and M~Teboulle.
\newblock A fast iterative shrinkage-thresholding algorithm for linear inverse
  problems.
\newblock \emph{SIAM Journal on Imaging Sciences}, 2\penalty0 (1):\penalty0
  183--202, 2009{\natexlab{b}}.

\bibitem[Butcher(1987)]{butcher1987}
JC~Butcher.
\newblock \emph{The Numerical Analysis of Ordinary Differential Equations:
  {R}unge-{K}utta and General Linear Methods}.
\newblock Wiley-Interscience, 1987.

\bibitem[Cand{\`e}s(2006)]{candes2006}
EJ~Cand{\`e}s.
\newblock Compressive sampling.
\newblock In \emph{Proceedings oh the International Congress of Mathematicians:
  Madrid, August 22-30, 2006: invited lectures}, pages 1433--1452, 2006.

\bibitem[Cand{\`e}s(2008)]{candes2008}
EJ~Cand{\`e}s.
\newblock The restricted isometry property and its implications for compressed
  sensing.
\newblock \emph{Comptes Rendus Mathematique}, 346\penalty0 (9):\penalty0
  589--592, 2008.

\bibitem[Cand{\`e}s and Tao(2005)]{candes2005}
EJ~Cand{\`e}s and T~Tao.
\newblock Decoding by linear programming.
\newblock \emph{IEEE Transactions on Information Theory}, 51\penalty0
  (12):\penalty0 4203--4215, 2005.

\bibitem[Catlin et~al.(2001)Catlin, Abkowitz, and Guttorp]{catlin2001}
SN~Catlin, JL~Abkowitz, and P~Guttorp.
\newblock Statistical inference in a two-compartment model for hematopoiesis.
\newblock \emph{Biometrics}, 57\penalty0 (2):\penalty0 546--553, 2001.

\bibitem[Cattamanchi et~al.(2006)Cattamanchi, Hopewell, Gonzalez, Osmond,
  Masae~Kawamura, Daley, and Jasmer]{cattamanchi2006}
A~Cattamanchi, PC~Hopewell, LC~Gonzalez, DH~Osmond, L~Masae~Kawamura, CL~Daley,
  and RM~Jasmer.
\newblock A 13-year molecular epidemiological analysis of tuberculosis in {S}an
  {F}rancisco.
\newblock \emph{The International Journal of Tuberculosis and Lung Disease},
  10\penalty0 (3):\penalty0 297--304, 2006.

\bibitem[Crawford et~al.(2014)Crawford, Minin, and Suchard]{crawford2013}
FW~Crawford, VN~Minin, and MA~Suchard.
\newblock Estimation for general birth-death processes.
\newblock \emph{Journal of the American Statistical Association}, 109\penalty0
  (506):\penalty0 730--747, 2014.

\bibitem[Donoho(2006)]{donoho2006}
DL~Donoho.
\newblock Compressed sensing.
\newblock \emph{IEEE Transactions on Information Theory}, 52\penalty0
  (4):\penalty0 1289--1306, 2006.

\bibitem[Doss et~al.(2013)Doss, Suchard, Holmes, Kato-Maeda, and
  Minin]{doss2013}
CR~Doss, Ma~Suchard, I~Holmes, MM~Kato-Maeda, and VN~Minin.
\newblock Fitting birth--death processes to panel data with applications to
  bacterial {DNA} fingerprinting.
\newblock \emph{The Annals of Applied Statistics}, 7\penalty0 (4):\penalty0
  2315--2335, 2013.

\bibitem[Doucet et~al.(2000)Doucet, Godsill, and Andrieu]{doucet2000}
A~Doucet, S~Godsill, and C~Andrieu.
\newblock On sequential {M}onte carlo sampling methods for {B}ayesian
  filtering.
\newblock \emph{Statistics and computing}, 10\penalty0 (3):\penalty0 197--208,
  2000.

\bibitem[El-Hay et~al.(2006)El-Hay, Friedman, Koller, and Kupferman]{elhay2006}
T~El-Hay, N~Friedman, D~Koller, and R~Kupferman.
\newblock Continuous time {M}arkov networks.
\newblock In \emph{Proceedings of the Twenty-second Conference on Uncertainty
  in AI (UAI)}, Boston, Massachussetts, July 2006.

\bibitem[Fong et~al.(2009)Fong, Guttorp, and Abkowitz]{fong2009}
Y~Fong, P~Guttorp, and J~Abkowitz.
\newblock Bayesian inference and model choice in a hidden stochastic
  two-compartment model of hematopoietic stem cell fate decisions.
\newblock \emph{The Annals of Applied Statistics}, 3\penalty0 (4):\penalty0
  1695--1709, 12 2009.

\bibitem[Golinelli et~al.(2006)Golinelli, Guttorp, and Abkowitz]{golinelli2006}
D~Golinelli, P~Guttorp, and JA~Abkowitz.
\newblock Bayesian inference in a hidden stochastic two-compartment model for
  feline hematopoiesis.
\newblock \emph{Mathematical Medicine and Biology}, 23\penalty0 (3):\penalty0
  153--172, 2006.

\bibitem[Grassmann(1977)]{grassmann1977}
WK~Grassmann.
\newblock Transient solutions in {M}arkovian queueing systems.
\newblock \emph{Computers \& Operations Research}, 4\penalty0 (1):\penalty0
  47--53, 1977.

\bibitem[Guttorp(1995)]{guttorp1995}
P~Guttorp.
\newblock \emph{Stochastic modeling of scientific data}.
\newblock CRC Press, 1995.

\bibitem[Hajiaghayi et~al.(2014)Hajiaghayi, Kirkpatrick, Wang, and
  Bouchard{-}C{\^{o}}t{\'{e}}]{hajiaghayi2014}
M~Hajiaghayi, B~Kirkpatrick, L~Wang, and A~Bouchard{-}C{\^{o}}t{\'{e}}.
\newblock Efficient continuous-time {M}arkov chain estimation.
\newblock In \emph{Proceedings of the 31th International Conference on Machine
  Learning, {ICML} 2014, Beijing, China, 21-26 June 2014}, pages 638--646,
  2014.

\bibitem[Lange(1982)]{lange1982}
K~Lange.
\newblock Calculation of the equilibrium distribution for a deleterious gene by
  the finite {F}ourier transform.
\newblock \emph{Biometrics}, 38\penalty0 (1):\penalty0 79--86, 1982.

\bibitem[Minin and Suchard(2008)]{minin2008}
VN~Minin and MA~Suchard.
\newblock Counting labeled transitions in continuous-time {M}arkov models of
  evolution.
\newblock \emph{Journal of Mathematical Biology}, 56\penalty0 (3):\penalty0
  391--412, 2008.

\bibitem[Nesterov(1983)]{nesterov1983}
Y~Nesterov.
\newblock A method of solving a convex programming problem with convergence
  rate {O}(1/k2).
\newblock In \emph{Soviet Mathematics Doklady}, volume~27, pages 372--376,
  1983.

\bibitem[Orkin and Zon(2008)]{orkin2008}
SH~Orkin and LI~Zon.
\newblock Hematopoiesis: An evolving paradigm for stem cell biology.
\newblock \emph{Cell}, 132\penalty0 (4):\penalty0 631--644, 2008.

\bibitem[Rao and Teh(2011)]{rao2011}
VA~Rao and YW~Teh.
\newblock Fast {MCMC} sampling for {M}arkov jump processes and continuous time
  {B}ayesian networks.
\newblock In \emph{Proceedings of the 27th International Conference on
  Uncertainty in Artificial Intelligence}. 2011.

\bibitem[Renshaw(2011)]{renshaw2011}
E~Renshaw.
\newblock \emph{Stochastic Population Processes: Analysis, Approximations,
  Simulations}.
\newblock Oxford University Press Oxford, UK, 2011.

\bibitem[Rosenberg et~al.(2003)Rosenberg, Tsolaki, and Tanaka]{rosenberg2003}
NA~Rosenberg, AG~Tsolaki, and MM~Tanaka.
\newblock Estimating change rates of genetic markers using serial samples:
  applications to the transposon {IS}\textit{6110} in \textit{{M}ycobacterium
  tuberculosis}.
\newblock \emph{Theoretical Population Biology}, 63\penalty0 (4):\penalty0
  347--363, 2003.

\bibitem[Shannon(2001)]{shannon2001}
CE~Shannon.
\newblock A mathematical theory of communication.
\newblock \emph{ACM SIGMOBILE Mobile Computing and Communications Review},
  5\penalty0 (1):\penalty0 3--55, 2001.

\bibitem[Tibshirani(1996)]{tibshirani1996}
R~Tibshirani.
\newblock Regression shrinkage and selection via the lasso.
\newblock \emph{Journal of the Royal Statistical Society. Series B
  (Methodological)}, pages 267--288, 1996.

\bibitem[Xu et~al.(2014)Xu, Guttorp, Kato-Maeda, and Minin]{xu2014}
J~Xu, P~Guttorp, MM~Kato-Maeda, and VN~Minin.
\newblock Likelihood-based inference for discretely observed birth-death-shift
  processes, with applications to evolution of mobile genetic elements.
\newblock \emph{ArXiv e-prints}, arXiv:1411.0031, 2014.

\end{thebibliography}

\setcounter{table}{0}
\renewcommand{\thetable}{C-\arabic{table}}
\renewcommand{\thefigure}{C-\arabic{figure}}

\renewcommand{\theequation}{A-\arabic{equation}}
\setcounter{equation}{0}

\section*{Supplement}

\subsection*{Discrete Fourier matrix}
The  $N$ by $N$ discrete Fourier transform matrix  $\mb{F}_N$ has entries $$\left\{ \mb{F}_N \right\}_{j,k} = \frac{1}{\sqrt{N}} (\omega)^{jk}$$ with $j,k = 0, 1, \ldots, N-1$ and $\omega = e^{i 2\pi /N}$, and as we mention in the main paper, the inverse Fourier transform matrix $\boldsymbol \psi$ is given by its conjugate transpose. The partial $M$ by $N$ IDFT matrices $\mb{A}$ necessary in Algorithm 1 is obtained by only computing and stacking a subset of $M$ random rows from $\boldsymbol \psi$.

\subsection*{Line search subroutine}
We select step sizes with a simple line search algorithm summarized in the pseudocode below that works by evaluating an easily computed upper bound $\hat f$ on the objective $f$:
\begin{equation}\label{eq:upper}
 \hat{f}_L(Z,Y) := f(Y) + \nabla f(Y)^T(Z - Y) + \frac{L}{2} || Z - Y||_2^2 . \end{equation}
 We follow Beck and Teboulle [2009], who provide further details.
In implementation, we select $L = .000005$ and $c = .5$, and reuse the gradient computed in \texttt{line-search} for step 10 of Algorithm 1 in the main paper.

\begin{algorithm}[h!]
   \caption{\texttt{line-search} procedure.}
   \label{alg:linesearch}
\begin{algorithmic}[1]
\STATE \textbf{Input:} initial step size $L$, shrinking factor $c$, matrices $Y_k, \nabla g(Y_k)$.

\STATE Set $Z = \text{softh}(Y_k - L \nabla g(Y_k))$
\WHILE {$g(Z) > \hat{f}_L(Z, Y_{k})$}
\STATE Update $L = c L$

\ENDWHILE
\RETURN $L_k = L$
\end{algorithmic}
\end{algorithm}

\subsection*{Derivation for hematopoiesis process PGF}
Given a two-type branching process defined by instantaneous rates $a_i(k,l)$, denote the following \textit{pseudo-generating} functions for $i = 1,2$:
\[ u_i(s_1,s_2) = \sum_k \sum_l a_i(k,l)s_1^k s_2^l \]

We may expand the probability generating functions in the following form:
\begin{align*}
\phi_{10}(t, s_1, s_2) &= E (s_1^{X_1(t)} s_2^{X_2(t)} | X_1(0) = 1, X_2(0) = 0)
\\	&= \sum_{k=0}^\infty \sum_{l=0}^\infty P_{(1,0), (k,l)} (t) s_1^k s_2^l 
\\ 	&= \sum_{k=0}^\infty \sum_{l=0}^\infty ( \mathbf{1}_{k=1, l = 0} + a_1(k,l) t + o(t) ) s_1^k s_2^l
\\ 	&= s_1 + u_1(s_1, s_2) t + o(t).
\end{align*}

Of course we have an analogous expression for $\phi_{01}(t, s_1, s_2)$ beginning with one particle of type 2 instead of type 1. For short, we will write $\phi_{10} := \phi_1, \phi_{01} := \phi_2$.

Thus we have the following relation between the functions $\phi$ and $u$:
\[\frac{d \phi_1}{dt} (t, s_1, s_2) |_{t=0} = u_1(s_1, s_2) \]
\[\frac{d \phi_2}{dt} (t, s_1, s_2) |_{t=0} = u_2(s_1, s_2) \]

To derive the backwards and forward equations, Chapman-Kolmogorov arguments yield the symmetric relations
\begin{align}
\phi_1(t+h, s_1, s_2) &= \phi_1(t, \phi_1(h, s_1, s_2), \phi_2(h, s_1, s_2))
\\ &= \phi_1(h, \phi_1(t, s_1, s_2), \phi_2(t, s_1, s_2))
\end{align}
First, we derive the backward equations by expanding around $t$ and applying $(2)$:
\begin{align*}
\phi_1(t+h, s_1, s_2) &= \phi_1(t, s_1, s_2) + \frac{d \phi_1}{dh}(t+h, s_1, s_2) | _{h=0} h + o(h) 
\\ &= \phi_1(t, s_1, s_2) + \frac{d \phi_1}{dh}(h, \phi_1(t, s_1, s_2), \phi_2(t, s_1, s_2) |_{h=0} h + o(h) 
\\ &= \phi_1(t,s_1,s_2) + u_1( \phi_1(t,s_1, s_2) , \phi_2(t, s_1, s_2) h + o(h) )
\end{align*}
Since an analogous argument applies for $\phi_2$, we arrive at the system
$$
\begin{cases}
\frac{d}{dt} \phi_1(t, s_1, s_2) = u_1( \phi_1(t, s_1, s_2), \phi_2(t, s_1, s_2) ) \\
\frac{d}{dt} \phi_2(t, s_1, s_2) = u_2( \phi_1(t, s_1, s_2), \phi_2(t, s_1, s_2) ) 
\end{cases}
$$
with initial conditions $\phi_1(0, s_1, s_2) = s_1, \phi_2(0, s_1, s_2) = s_2$.

Recall the rates defining the two-compartment hematopoiesis model are given by
\begin{align*}
a_1(2,0) &= \rho && a_1(0,1) = \nu && a_1(1,0) = -(\rho + \nu) \\
a_2(0,0) &= \mu && a_2(0,1) = -\mu
\end{align*}
Thus, the pseudo-generating functions are 
\[ u_1(s_1, s_2) = \rho s_1^2 + \nu s_2 - (\rho + \nu) s_1 \]
\[ u_2(s_1, s_2) = \mu - \mu s_2 = \mu(1 - s_2) \]
Plugging into the backward equations, we obtain 
\[\frac{d}{dt} \phi_1(t,s_1,s_2) = \rho \phi_1^2(t, s_1, s_2) + \nu \phi_2(t, s_1, s_2) - (\rho+ \nu) \phi_1(t, s_1, s_2) \]
and 
\[ \frac{d}{dt} \phi_2(t,s_1, s_2) = \mu - \mu \phi_2(t,s_1,s_2). \]
The $\phi_2$ differential equation corresponds to a pure death process and is immediately solvable: suppressing the arguments of $\phi_2$ for notational convenience, we obtain
\begin{align*}
 \frac{d}{dt} \phi_2 &= \mu - \mu \phi_2
 \\ \frac{d}{dt} \phi_2( \frac{1}{1 - \phi_2}) &= \mu
 \\ \ln (1 - \phi_2) &= -\mu t + C 
 \\ \phi_2 &= 1 - \exp(-\mu t + C)
\end{align*}
Pluggin in $\phi_2(0, s_1, s_2) = s_2$, we obtain $C = \ln(1-s_2)$, and we arrive at 
\begin{equation}\label{phi2}
\phi_2(t, s_1, s_2) = 1 + (s_2 - 1)\exp(-\mu t) 
\end{equation}

Plugging this solution into the other backward equation, we obtain 
\begin{equation}\label{phi1}
\frac{d}{dt} \phi_1(t, s_1, s_2) = \rho \phi_1^2(t, s_1, s_2) - (\rho + \nu) \phi_1(t,s_1, s_2) + \nu(1 + (s_2 - 1)\exp(-\mu t) ) \end{equation}

This ordinary differential equation can be solved numerically given rates and values for the three arguments, allowing computation of $\phi_{i,j} = \phi_1^i \phi_2^j$ which holds by particle independence. 

\subsection*{BDS model diagram}
 The branching process components $\mb{X}(t)= (x_{old}, x_{new})$ represent the number of originally occupied and newly occupied sites at the end of each observation interval. As an example, assume six particles (transposons) are present initially at time $t_0$,  and a shift and a birth occur before the first observation $t_1$, and a death occurs before a second observation at $t_2$. When considering the first observation interval $[t_0,t_1)$, we have $ \left\{ \mb{X}(t_0) = (6,0), \mb{X}(t_1)=  (5,2) \right\}$. When computing the next transition probability over $[t_1, t_2)$, we now have $\left\{ \mb{X}(t_1) = (7,0), \mb{X}(t_2) = (6,0) \right\}$, since all seven of the particles at $t_1$, now the left endpoint of the observation interval, now become the initial population. Even with data over time, this seeming inconsistency at the endpoints does not become a problem because transition probability computations occur separately over disjoint observation intervals. See Xu et al. [2014] for further details.
 
\begin{figure}
\centering
\includegraphics[width = .45\paperwidth]{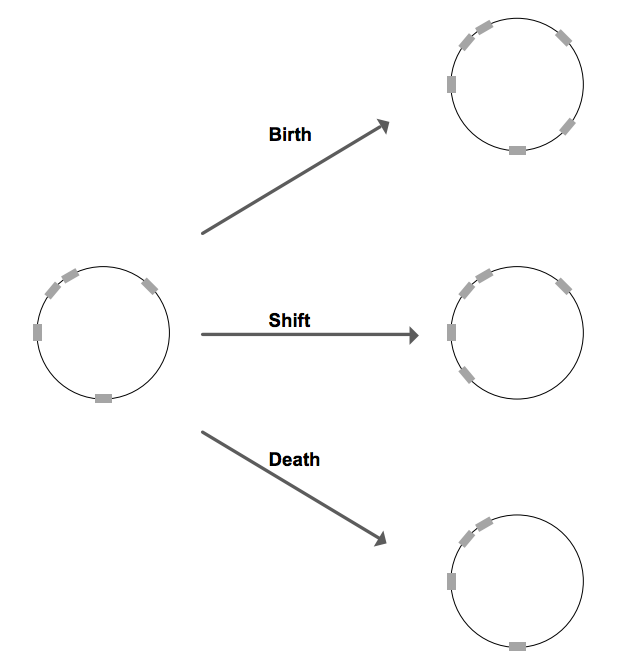}
\caption{ Illustration of the three types of transposition---birth, death, shift---along a genome, represented by circles [Rosenberg et al., 2003]. Transposons are depicted by rectangles occupying locations along the circles/genomes. 
On the right set of diagrams, a birth event keeps the number of type 1 particles intact and increments the number of type 2 particles by one, a death event changes the number of type 1 particles from five to four and keeps the number of type 2 particles at zero, and finally a shift event decreases the number of type 1 particles by one and increases the number of type 2 particles by one.} 
\label{fig:BDS}
\end{figure}

\end{document}